\documentclass[aps,prl,twocolumn,showpacs,superscriptaddress,groupedaddress,jmp,amsmath,amssymb,reprint]{revtex4-1} 
\usepackage{graphicx}
\DeclareGraphicsExtensions{.png,.jpg}
\usepackage{dcolumn}
\usepackage{bm}
\usepackage{natbib}
\usepackage{nicefrac}
\usepackage{amsfonts}
\usepackage{multirow} 
\usepackage{hyperref}
\usepackage{CJK}

\begin{document}
	\title{Nonlinear Flexoelectricity in Non-centrosymmetric Crystals}
	\author{Kanghyun Chu}
	\email{kanghyunchu@kaist.ac.kr}
	\affiliation{Department of Physics, KAIST, Daejeon, 305-701, Republic of Korea}
	\author{Chan-Ho Yang}
	\email{chyang@kaist.ac.kr}
	\affiliation{Department of Physics, KAIST, Daejeon, 305-701, Republic of Korea}
%	\date{\today}
\begin{abstract}
	We analytically derive the elastic, dielectric, piezoelectric, and the flexoelectric phenomenological coefficients as functions of microscopic model parameters such as ionic positions and spring constants in the two-dimensional square-lattice model with rock-salt-type ionic arrangement.
	Monte-Carlo simulation reveals that a difference in the given elastic constants of the diagonal springs, each of which connects the same cations or anions, is responsible for the linear flexoelectric effect in the model. 
	We show the quadratic flexoelectric effect is present only in non-centrosymmetric systems and it can overwhelm the linear effect in feasibly large strain gradients. 
\end{abstract}
		% 77.65.-j 	Piezoelectricity and electromechanical effects
		% 77.84.-s 	Dielectric, piezoelectric, ferroelectric, and antiferroelectric materials
		% 78.20.Bh  Theory, models, and numerical simulation
% 	\keywords{flexoelectric, strain gradient}
	\maketitle

	Flexoelectricity, the inducement of an electric polarization by strain gradients, is an electromechanical phenomenon inherent in all dielectric materials in any space group \cite{Kogan1963,Tagantsev1985,Tagantsev1986,Zubko2013,Mao2014}.
	Despite its ubiquity, the study of flexoelectricity was mainly focused on soft materials and liquid crystals \cite{Meyer1969,Helfrich1971,Osipov1983}.
	The flexoelectric effect in rigid materials was considered insignificant compared to other electromechanical phenomena such as piezoelectricity, because a large strain gradient is hard to attain in macroscopic systems without fracturing or cracking. 
	However, the recent advancement of nanoscale technology enables us to manipulate atomic scale systems such as strain relaxation in misfit strained epitaxial thin films \cite{Catalan2005,Lee2011,Jeon2013}, domain walls and interfaces \cite{Catalan2011,Borisevich2012,Chu2015}, and tip-induced inhomogeneous mechanical deformation \cite{Lu2012}. 
	Observation of a giant strain gradient in the range of $10^5 \! \sim \! 10^7 \text{ m}^{-1}$ is not astonishing any more in nanoscale research \cite{Lee2011,Jeon2013,Nguyen2013a,Zeches13112009}.

	Currently the relation between ferroelectricity and strain gradients becomes an important topic of research in dielectrics.
	The flexoelectric coefficients of some oxide materials have been carefully determined by experiments \cite{Ma2005,Cross2006,Zubko2007}. 
	The development of calculation methods and the simulation studies have also provided a deep understanding of the flexoelectricity \cite{Kalinin2008,Hong2011,Xu2013,Stengel2014,Salje2016}.
	But still, the flexoelectric effect under a huge strain gradient, whereby a nonlinear response arises, is little studied.
	Considering the crystal symmetry, the second order flexoelectric effect which is described by seventh (odd) order tensor becomes non-zero if the system does not possess an inversion center such as piezoelectric or ferroelectric materials \cite{Chu2015,Cronin1993,Naumov2009}.

	In this Letter, we elucidate the microscopic origin of the flexoelectric effect and evaluate the relative strength of the quadratic and linear flexoelectric effects. 
	Starting from the analytic derivation of the electromechanical properties in a one-dimensional ionic chain model, we expand our discussion on the flexoelectricity into two-dimensional systems. 

\begin{figure}[b]
	\includegraphics{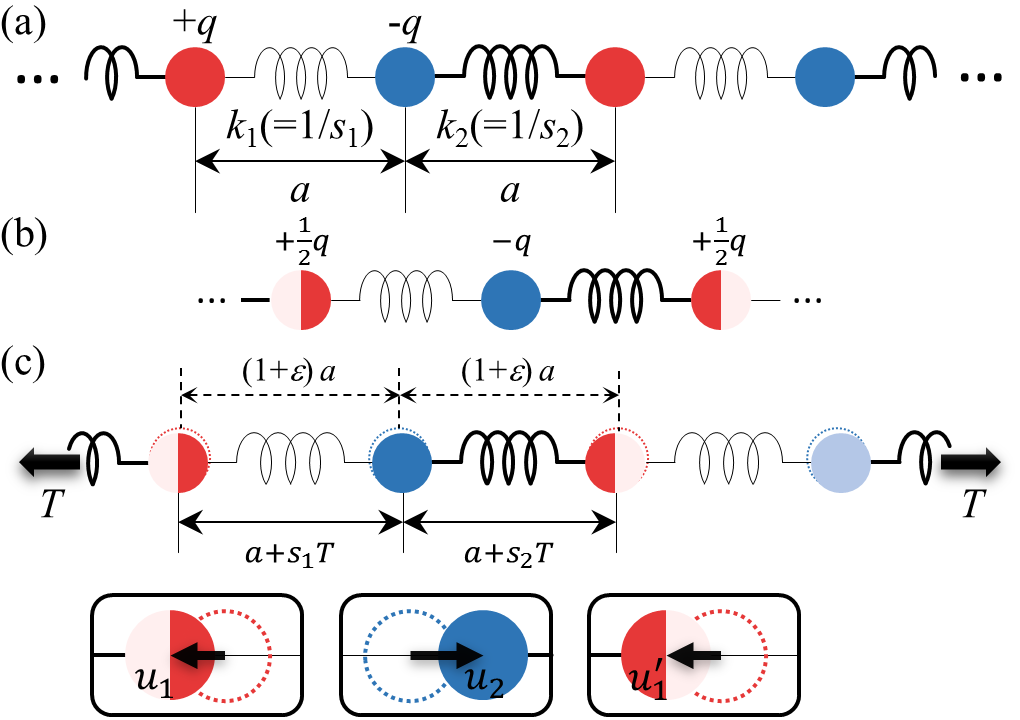}
	\caption{One-dimensional ionic chain model. (a) The unperturbed state. (b) A dipole-free unit cell. (c) Intrinsic piezoelectric effect. The $u_1$, $u_2$ and $u_1'$ are internal strains. The figure shows a situation where the system of $k_2>k_1$ is under a tensile force. } 
	\label{fig1}
\end{figure}

	First, we introduce how to extract the intrinsic piezoelectric effect with excluding the surface piezoelectric effect based on a one-dimensional microscopic model in a pedagogical way before beginning a more complex description regarding the strain gradient effects in a higher dimension. 
	It is also essential in terms of the fact that the analytic form of the piezoelectric polarization is a part of the induced polarization in the case of a strain gradient.
	Our starting ionic chain model is composed of two parts: point masses (with alternating electric charges $\pm q$) and harmonic massless springs (characterized by elastic constants $k_{1, 2}$ or their inverse values called elastic compliances $s_{1, 2}$) as shown in the Figure \ref{fig1}(a).
	We assume the alternating positive and negative ions are equally spaced by a distance $a$ at no external perturbations indicating the lattice parameter is $2a$. 
	The basic mechanical and dielectric responses of the model are given by $\varepsilon=\frac{1}{2a}(s_1+s_2)T$ and $\Delta p_{\text{u.c.}}=\frac{1}{4}q^2(s_1+s_2)E$ respectively \cite{supple}, where $\varepsilon$ is strain, $T$ is applied tensional force, $\Delta p_{\text{u.c.}}$ is induced dipole moment per unit cell, and $E$ is external electric field. 
	If we need to break the inversion symmetry, we can take different values of the elastic constants ($k_1 \neq k_2$) or/and choose different inter-ionic spacing.
	In order to clarify the intrinsic piezoelectric effect, it is convenient to take a dipole-free unit cell as described in the Figure \ref{fig1}(b) because the extrinsic piezoelectricity is reduced to a surface charge effect and its contribution automatically drops from the bulk calculation as shown in the following equation \cite{Tagantsev1991}:
\begin{equation}
\begin{aligned}
	p_{\text{u.c.}} &=\sum_{\alpha}q_{\alpha}X_{\alpha}(\varepsilon) =\sum_{\alpha}q_{\alpha} \big( (1+\varepsilon)X_{\alpha}^0+u_{\alpha}(\varepsilon) \big) \\
					&=\sum_{\alpha}q_{\alpha}u_{\alpha}(\varepsilon),
\end{aligned}
\end{equation}
	where $p_{\text{u.c.}}$ is the dipole moment per unit cell and $\alpha$ stands for the ionic index within a unit cell. 
	$X_{\alpha}$ represents the position of $\alpha$-th ion when the strain $\varepsilon$ is applied, while
	$X_{\alpha}^0$ represents the original coordinate, \textit{i.e.} $X_{\alpha}(\varepsilon=0) = X_{\alpha}^0$.
	$u_{\alpha}$ represents the internal strain imposed on $\alpha$-th ion, and it indicates a displacement from the point into which a linear scaling transformation by factor $1+\varepsilon$ transforms $X_{\alpha}^0$ \cite{Tagantsev1986}.
	The dipole-free condition guarantees the equality $\sum_{\alpha}q_{\alpha}X_{\alpha}^0 = 0$.
	
	For given $T$, $\varepsilon$ is $\frac{1}{2a}(s_1+s_2)T$ since the length change of each spring is $s_{1,2}T$.
	The difference between the internal strains $u_2-u_1$ is calculated as $\frac{1}{2}(s_2-s_1)T$.
	We note $u_1$ is equal to $u_1'$ because the discrete translational symmetry is conserved. 
	So, the induced piezoelectric dipole moment per unit cell $p_{\text{u.c.}}$ is written as
\begin{equation}
\begin{aligned}
	p_{\text{u.c.}} &\!=\! \frac{q}{2} u_1 \! - \! q u_2 \! +\! \frac{q}{2} u_1' \!=\! q(u_1-u_2) \!=\! \frac{q}{2}(s_2-s_1)T \\
				\Big(&=qa\frac{s_2-s_1}{s_1+s_2} \varepsilon\Big).
\end{aligned}
\end{equation}
	Thus, the piezoelectric coefficient, dipole moment per unit stress normalized by system size, is $\frac{p_{\text{u.c.}}}{2aT} = \frac{q}{4a}(s_2-s_1)$. 
	We can identify that the converse piezoelectric coefficient - induced strain per unit electric field - gives the same result, considering a uniform electric field ($E$) provokes a strain as $\varepsilon(E)=\frac{1}{2a}\Big(\frac{1}{2}(s_2-s_1)qE\Big)$ \cite{supple}.

\begin{figure}[t]
	\includegraphics{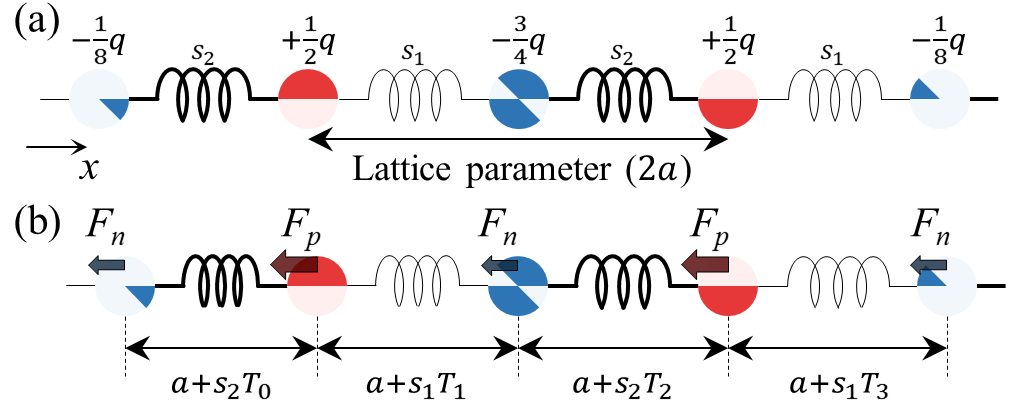}
	\caption{One-dimensional flexoelectric effect. (a) Dipole- and quadrupole-free unit cell (b) Strain gradient is driven by external forces acting on cations and anions, which are denoted by $F_p$ and $F_n$ respectively. For a positive strain gradient of $\frac{\partial \varepsilon}{\partial x}$, the $F_p+F_n$ is negative. The intrinsic flexoelectric property is determined by the ratio between $F_p$ and $F_n$.}
	\label{fig2}
\end{figure}
	With this in mind, we speculate effects on electric polarization by strain gradients in the one-dimension system.
	For the convenience in deriving the flexoelectric coefficient, the unit cell is taken so that it has no electric quadrupole as well as no dipole moment, \textit{i.e.}, satisfying the relations: $\sum_{\alpha} q_{\alpha}X_{\alpha}=0 $ and $\sum_{\alpha}^{} q_{\alpha}X_{\alpha}^2=0$.
	Similar to the fact that the dipole-free unit cell is useful in ruling out the extrinsic piezoelectric effect, the virtual unit cell shown in the Figure \ref{fig2}(a) automatically excludes the surface flexoelectric effect arising from a non-zero quadrupole moment of the system \cite{Tagantsev1986}.
	The unit cell contains partial ions with fractional charges and masses.
	Although the size of unit cell is larger than the real lattice parameter ($2a$), the repetition of the cells at the interval of $2a$ constructs the original lattice by permission of the overlap.
	But, additional surface charges in finite systems will be necessarily introduced to compensate absence of the missing partial charges on the terminations. 
	This is a mathematical trick to nullify the pole moments up to the second order term and it has merits in studying bulk properties by explicitly separating them from extrinsic effects. 

	Provided the system has a homogeneous strain gradient of $\acute{\varepsilon}$ as shown in the Figure \ref{fig2}(b), it is necessarily involved in a tension gradient. 
	They are related each other by a mechanical coefficient, \textit{i.e.}
\begin{equation}
	\frac{\mathit{\Delta}\varepsilon}{\mathit{\Delta}x} (\equiv\acute{\varepsilon})=\frac{1}{2a} (s_1+s_2) \frac{\mathit{\Delta}T}{\mathit{\Delta}x},
\end{equation}
	where the $x$ is the spatial coordinate. 
	From a microscopic point of view, the phenomenological strain gradient increases the tension force exerted on the next-nearest-neighboring spring, which has the same elastic constant in the next unit cell, by $2a\frac{\mathit{\Delta}T}{\mathit{\Delta}x}$ as compared with that of the original unit cell.
	However, the phenomenological gradient cannot uniquely determine the relative intra-unit-cell deformations in one-dimensional cases. 
	We have an extra degree of freedom regarding the relative strength of so called body forces exerted on the positive and negative ions ($F_p$ and $F_n$).  
	They are responsible for increments between the tension forces of two neighboring springs. 

	The tension force on each spring, \textit{i.e.}, $T_0, T_1, T_2$ and $T_3$ assigned from the leftmost spring, should satisfy the conditions: $T_1 =T_0 - F_p$, $T_2 =T_1 - F_n$, and $T_3 =T_2 - F_p$ for the force balance at each ion.
	The positions of ions are also obtained in a recursive manner as $X_1=X_0 + a + s_2 T_0$,$X_2 =X_1 + a + s_1 T_1$, $X_3 =X_2 + a + s_2 T_2$, and $X_4 =X_3 + a + s_1 T_3$.
	The induced dipole in the cell is obtained finally as the following:
\begin{equation}
\begin{aligned}
	\sum_{\alpha=0}^{4}q_{\alpha}X_{\alpha}\!&=\!\frac{q}{8}(\!s_1\!+\!s_2\!)(\! F_p \! - \! F_n \! )\!+\!\frac{q}{8}(\!s_2\!-\!s_1\!)\Big(\sum_{i=0}^{3}T_i \Big) \\
	&=\frac{1}{2}qa^2\frac{F_p-F_n}{F_p+F_n}\acute{\varepsilon} + \frac{q}{2}(s_2-s_1)\bar{T}.
\end{aligned}
\end{equation} 
	Note that strain gradient (and the corresponding macroscopic tension gradient) is proportional to the $F_p+F_n$.
	The first term in the last equation is proportional to the given strain gradient $\acute{\varepsilon}$ and thus the proportional coefficient corresponds to the intrinsic flexoelectric effect. 
	The second term depending on the inversion symmetry breaking ($s_2-s_1$) is due to the intrinsic piezoelectric effect and it has a position-dependence as such mean tension of unit cell ($\bar{T}$) does.
	From the derivation, we can lead to the conclusion that the dimensionless $\frac{F_p-F_n}{F_p+F_n}(\equiv f)$ is the origin of the linear flexoelectric effect. 
	The $f$ seems to be arbitrarily chosen in one-dimensional cases although this ambiguity is removed in higher dimensions as will be addressed below. 
	The value can be temporally specified, as an characteristic of the one-dimensional model, by a sort of bonding property unrelated to the inversion symmetry breaking or charges of constituent atoms in cases of a uniform external field.

\begin{figure}[t] 
	\includegraphics{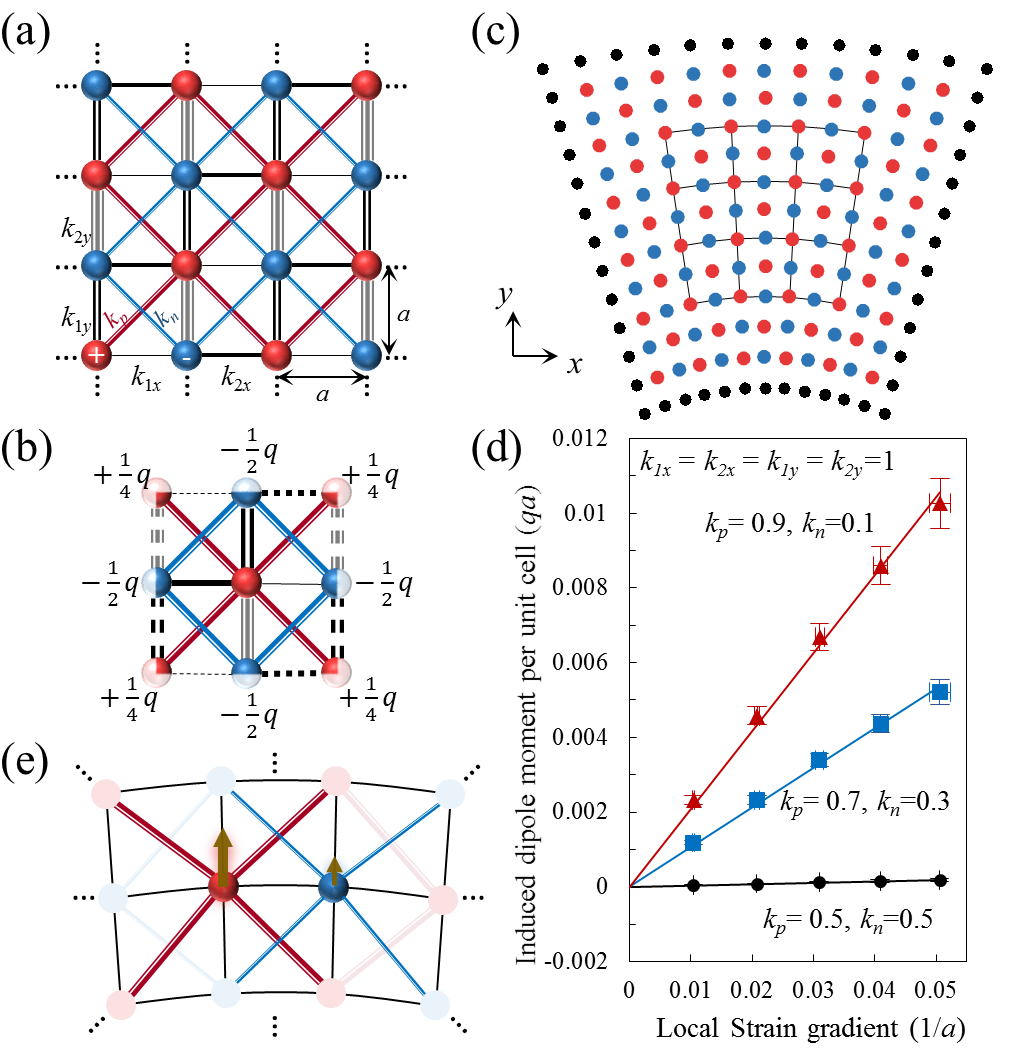}
	\caption{Two dimensional ionic chain network model. 
		(a) Model scheme. 
		(b) Pole-free cell. 
		(c) Typical relaxed configuration of the strain gradient of $0.05/a$. Solid black dots are fixed sites. The black guide lines illuminate the 3-by-3 pole-free cells in use. 
		(d) Induced dipole moment with respect to applied strain gradient. 
		(e) Schematic diagram of linear flexoelectricity.}
	\label{fig3}
\end{figure} 
\begin{figure}[t]
	\includegraphics{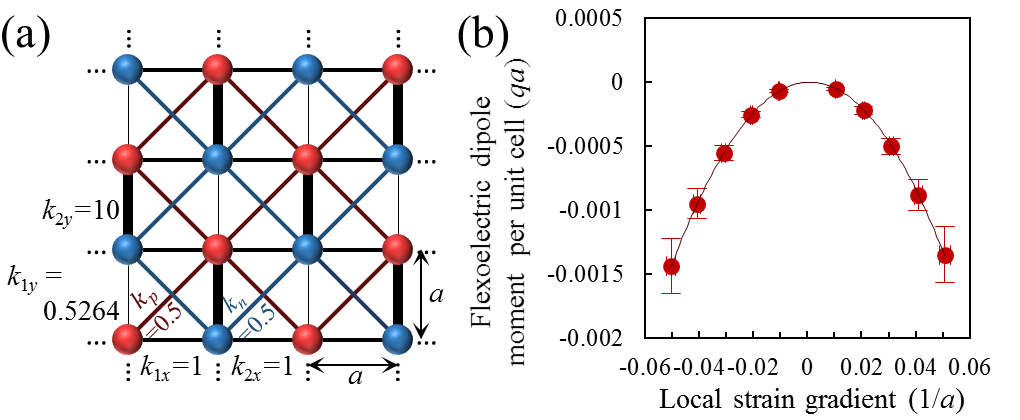}
	\caption{Quadratic flexoelectric effect in a non-centrosymmetric system. 
		(a) An ionic model with no inversion symmetry as a consequence of the difference in $k_{1y}$ and $k_{2y}$. 
		(b) Dipole moment per unit cell flexoelectrically induced by the local strain gradient. } 
	\label{fig4}
\end{figure}

%% 2D simulation study part 
	We have performed the simulation study based on a two-dimensional ionic model. 
	It consists of positive and negative ions and elastic springs as shown in the Figure \ref{fig3}(a).
	There are six types of springs: two along $x$-axis $k_{1x}$, $k_{2x} $, other two along $y$-axis $k_{1y}$, $k_{2y}$, and the last two connecting between ions of the same charges $k_p$, $k_n$ along diagonals. 
	It catches the essence of nearest- and next nearest-neighbor interactions in a two-dimensional rock salt structure which is an easily addressable crystalline system. 

	We introduce a pole-free doubled unit cell as displayed in the Figure \ref{fig3}(b). 
	It is convenient to describe the local states based on the cell, because the extrinsic piezoelectric and flexoelectric effects are automatically eliminated in the bulk calculation, and all extrinsic effects are converted into the problems related to surface charges and surface dipoles.  

	The transverse strain gradient of $\frac{\partial \varepsilon_{xx}}{\partial y}$ is applied to the model system composed of 5-by-5 pole-free cells (its size is $2a \times 2a$) as shown in the Figure \ref{fig3}(c).
	The outermost black dots are fixed sites \cite{supple}. 
	Only the 3-by-3 pole free cells out of 25 cells are included in the property evaluation to rule out possible clamping effects arising from the constraint edge sites. 
	All the positions of ions except for the edge sites are relaxed by the Monte Carlo algorithm until the total elastic energy stored in the springs is minimized.

	In result, the conventional linear flexoelectric effect was numerically reproduced and it could be manipulated by tuning the diagonal springs.
	The model systems were designed to be centrosymmetric by setting all springs lying along the $x$- and $y$-axes to be identical to prohibit any intervention of the intrinsic piezoelectric effect.
	While tuning the $k_p$ and $k_n$, the sum $k_p+k_n$ was kept constant to equalize the elastic property of each model system.
	The Figure \ref{fig3}(d) shows the induced flexoelectric dipole moment as a function of the transverse strain gradient for selected $k_p$, $k_n$ conditions. 
	Each data point and error bar represent the mean and standard deviation of the local transverse strain gradient and the local $y$-dipole moment harvested from each relaxed configuration containing the nine pole-free cells. 
	The linear relationship between the strain gradient and the induced dipole moment is found implying the conventional linear flexoelectricity.

	The lesson from the analytic derivation of the flexoelectric coefficient in the one-dimensional ionic chain is that the linear flexoelectric effect increases as the $f$ gets larger.
	As shown in the Figure \ref{fig3}(e), the transverse strain gradient squeezes the bottom layer compressing the diagonal springs in below while it expands the top layer stretching the diagonal springs in above. 
	If $k_p$ is larger than $k_n$, the deformation generates a larger tugging force on the positive ion than that on the negative one.
	As a consequence, the positive ion shifts upward more than the negative ion resulting in a positive flexoelectric polarization.

	Furthermore, the nonlinear flexoelectric effect arising in non-centrosymmetric systems is simulated.
	The inversion symmetry is broken along the $y$-axis by discriminating $k_{1y}$ and $k_{2y}$ (Fig. \ref{fig4}(a)).
	A compliance matching condition $s_{1x}+s_{2x} = s_{1y}+s_{2y}$ is held when modifying the model parameters in order to make the elastic property along the $x$- and $y$-axes equivalent. 
	In addition, the linear flexoelectric response is eliminated by putting $k_p = k_n$ to focus on the second order effect.
	It is noteworthy mentioning that, when $k_{1x}=k_{2x}=k_{1y}=k_{2y}$, the system belongs to the two-dimensional space group of \textit{p4mm}; if $k_{1x}=k_{2x}$, $k_{1y} \neq k_{2y}$, the symmetry is reduced to \textit{cm}.
	Since the system also has piezoelectricity, separating the piezoelectric contribution from the raw simulated data of the induced dipole moment is required to extract the pure flexoelectric effect.
	The local piezoelectric dipole moment was able to be estimated by using the piezoelectric coefficients and the local strain states \cite{supple}.
	The Figure \ref{fig4}(b) exhibits the flexoelectric dipole moment per unit cell obtained with varying the applied strain gradient in the non-centrosymmetric system, indicating the manifestation of a quadratic flexoelectric effect.
%%%  single column figure
\begin{figure}[t] 
	\includegraphics{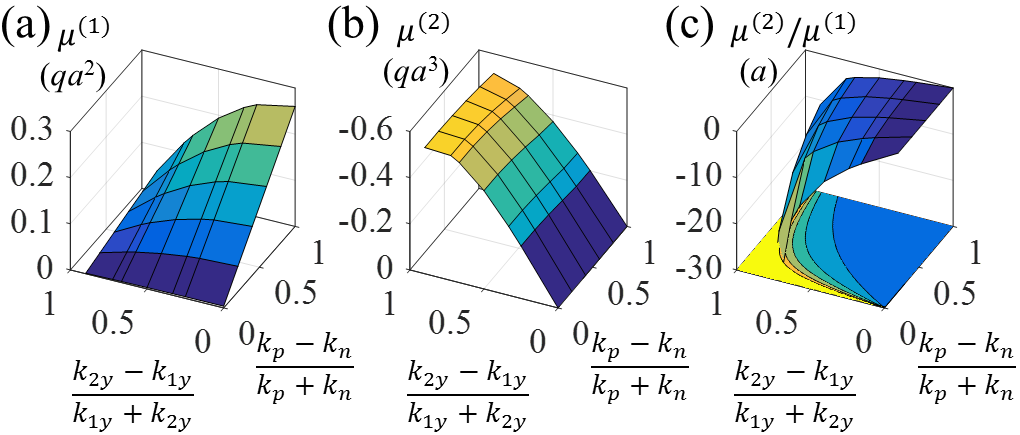}
	\caption{Flexoelectric coefficient maps. 
		(a) Linear flexoelectric coefficient $\mu^{(1)}$. 
		(b) Quadratic flexoelectric coefficient  $\mu^{(2)}$. 
		(c) Ratio of the quadratic coefficient to the linear coefficient.} 
	\label{fig5}
\end{figure}

	Finally, we systematically investigate the flexoelectric coefficients with respect to model parameters.
	The coefficients $\mu^{(1)}$ and $\mu^{(2)}$ can be determined by fitting the strain-gradient dependent data to the phenomenological relation: $p_{\text{u.c.}}^{\text{flexo}} = \mu^{(1)}\acute{\varepsilon} + \mu^{(2)}\acute{\varepsilon}^2$, where $p_{\text{u.c.}}^{\text{flexo}}$ is the induced flexoelectric dipole moment per pole-free cell, and $\acute{\varepsilon}$ is the transverse strain gradient.
	Fixing the spring constants along the $x$-axis ($k_{1x}=k_{2x}=1$), the spring constants along the $y$-axis and the diagonal axes were varied.
	At that time, the macroscopic elastic properties of systems were equalized by keeping the constraints: $k_p+k_n=1$ and $s_{1y}+s_{2y}=2$. 
	Each flexoelectric coefficient is plotted as a function of two dimensionless parameters of $\frac{k_p-k_n}{k_p+k_n}$ and $\frac{k_{2y}-k_{1y}}{k_{1y}+k_{2y}}$ which are related to the difference in the interaction strengths among the same kind of ions and the degree of inversion symmetry breaking, respectively.
	Interestingly, $\mu^{(1)}$ decreases as the inversion symmetry breaking gets severe (Fig. \ref{fig5}(a)).
	Meantime, $\mu^{(2)}$ is hardly affected by the first parameter (Fig. \ref{fig5}(b)).
	The plot of the ratio of the second order flexoelectric coefficient over the first order one (Fig. \ref{fig5}(c)) enables the estimation of the critical strain gradient above which the quadratic flexoelectric effect exceeds the linear effect.
	In cases that the linear flexoelectricity is tuned to diminish, the coefficient ratio reaches tens of ionic spacing $a$. 
	Considering the typical atomic distance of a few angstroms, the critical strain gradient is $\sim\!10^7\text{ m}^{-1}$, and it is a feasible value that is reported at the morphotropic phase interfaces \cite{Zeches13112009,Chu2015}.

%% discussion
	Our model system can be extended in various ways. 
	It is possible to consider a more complex unit cell that contains multiple ionic basis and extended neighbor interactions at the expense of rapidly increasing calculation complexity dealing with possible combinations of ions in a unit cell and its neighbors. 
	On the other hand, to handle a system made of a single kind of atoms, it can be treated as a pseudo binary ion system by introducing imaginary ions that stands for the center of electron cloud \cite{Marzari1997}.
	Adopting anharmonic springs allows the model system to represent the effects of electrostriction and second order piezoelectricity \cite{Sundar1992,Newnham1997,Bester2006,Grimmer2007}.

	The magnitudes of charges can be considered as the formal charges of the ions in a na\"{i}ve sense. 
	One may also interpret them as the Born effective charges taking electron redistribution into account \cite{Resta1994} and the spring constants as the Born-von Karman force constants \cite{Born1954} beyond the classical picture. 
	On the assumption that the charge is ten times larger than the elementary charge of electron and $a$ is a few $\text{\AA}$, the simulated result for the linear flexoelectric coefficient corresponds to $\sim\!1\text{ nC/m}$  which is comparable with the measured quantity in SrTiO$_3$ \cite{Zubko2007}.
	This theoretical approach opens the door to a quantitative understanding of the electromechanical properties in crystals.

	This work was supported by the National Research Foundation (NRF) Grant funded by the Korean Government (NRF-2013S1A2A2035418 and NRF-2014R1A2A2A01005979) and the NRF via the Center for Quantum Coherence in Condensed Matter (2016R1A5A1008184) and the Global Frontier R\&D Program on Center for Hybrid Interface Materials (2013M3A6B1078872).

\bibliography{Ionic_Chain_model_ver10.bib}

\end{document}